\documentclass[preprint,12pt]{elsarticle}
\journal{Nuclear Physics A}
\begin{document}
\begin{frontmatter}
\title{Theory of the electromagnetic production of hyperons \footnote{Presented 
at the 11th International Conference on Hypernuclear and Strange Particle 
Physics, 1 - 5 October, 2012, Barcelona.}}
\author{Petr Byd\v{z}ovsk\'y and Dalibor Skoupil}
\address{Nuclear Physics Institute of the ASCR, 
\v{R}e\v{z}, 250 68, Czech Republic}
\begin{abstract}
Some basic properties of isobar models are discussed using 
the Saclay-Lyon, Kaon-MAID, and H2 models and a comparison 
of their predictions with experimental data is given for photo- 
and electroproduction of $K^+$ on the proton and for photoproduction 
of $K^0$ on the deuteron in specific kinematical regions.
Results of the isobar models are also compared with the Regge-isobar 
hybrid model, Regge-plus-resonance.
\end{abstract}
\end{frontmatter}

\section{Introduction}

Production of the $\Lambda$ and $\Sigma$ hyperons on nucleons 
and nuclei induced by the electron beam provides completing 
information about properties of baryons and their behavior 
in nuclei. Besides the study of the reaction mechanism, a correct 
description of the elementary production on nucleons is important  
for minimizing uncertainties in calculations of the excitation 
spectra for electroproduction of hypernuclei~\cite{HProd,ByMa}.

There are various methods of description of the elementary 
production process. Among them the single-channel description 
based on an effective Lagrangian considering only hadronic 
degrees of freedom is of special importance because 
the corresponding isobar model can be easily utilized in the 
calculations of hypernucleus electroproduction~\cite{HProd}. 
Analyses of data with the isobar model provide information 
about properties of nucleon resonances and the existence of 
``missing resonances'' predicted by the quark model but not 
observed in other processes~\cite{MaBe,HYP03}.

In the hadrodynamical approach, several production channels are   
coupled by the final-state meson-baryon interaction and should 
be treated simultaneously. In this coupled-channel 
approach~\cite{CouCh}, rescattering effects in the meson-baryon 
system in intermediate states can be included.  
Considerable simplification originates in neglecting the 
rescattering effects in the formalism assuming that these 
effects are included to some extent by means of effective values 
of the coupling constants fitted to experimental data. 
This simplifying assumption was adopted in many isobar models, 
e.g., Saclay-Lyon (SL)~\cite{SL}, Kaon-MAID (KM)~\cite{MaBe,KM}, 
and Gent~\cite{Jan01}.

Another approach to description of the process, suited also for 
energies above the nucleon-resonance region up to 
$E_{\gamma}^{lab} \approx 16$ GeV, is the hybrid Regge-plus-resonance 
model~\cite{RPR,Lesley,LesleyPhD} (RPR). 
This model combines the Regge model~\cite{Regge}, appropriate 
to description above the resonance region ($E_{\gamma}^{lab}>4$ GeV), 
with elements of the isobar model eligible for the lower-energy 
region. 

In the quark models for photoproduction of kaons~\cite{SagQM}, resonances 
are implicitly included as excited states and therefore a number of 
free parameters is relatively small.
Another asset of this approach is a natural description of a hadron 
internal structure which have to be modeled phenomenologically via form 
factors in the isobar models.
However, the quark models for the electromagnetic production of 
kaons are too complicated for their further use in the 
calculations of hypernucleus electroproduction. 

\section{Isobar and Regge-plus-resonance models}  
In the isobar model the amplitude is constructed as a sum of 
the tree-level Feynman diagrams which can be divided into the 
nonresonant and resonant contributions. The former consists of 
the Born terms and exchanges of kaon (t-channel) and hyperon 
(u-channel) resonances. The latter is modeled by exchanges of 
nucleon resonances in s-channel. 
The problem of the isobar model for kaon photoproduction is a too 
large contribution from the Born terms which has to be reduced 
assuming some mechanism~\cite{Jan01}. 
One possibility is to include several hyperon resonances 
which counterbalance the Born contribution~\cite{SL,SLA}. 
Another way is to assume hadronic form factors (hffs) 
in the strong (baryon-meson-baryon) vertices~\cite{MaBe,KM} 
which suppress the Born terms very strongly. In the Gent 
isobar model a combination of both mechanisms is used~\cite{Jan01}.
Besides a reduction of the Born terms the hffs can model 
an internal structure of hadrons in the strong vertices 
neglected in the effective Lagrangian. 
The form factors are included by a gauge-invariant 
technique~\cite{DW} assuming the dipole~\cite{KM,Jan01}, 
Gaussian~\cite{RPR} or multidipole-Gauss~\cite{Lesley,LesleyPhD} 
types. 
These various methods of reduction of the Born terms influence 
strongly a dynamics of the isobar model. The problem of 
the large Born contributions is avoided in the RPR approach.   

Here we will discuss results of the KM, Saclay-Lyon A (SLA)~\cite{SLA}, 
and H2~\cite{HYP03} models. These models include the Born diagrams 
and contributions from exchanges of the $K^*(890)$ and $K_1(1270)$ 
resonances. The main coupling constants, $g_{KN\Lambda}$ and 
$g_{KN\Sigma}$, fulfill the limits of 20\% broken SU(3) 
symmetry~\cite{SL}.   
These models differ in a selection of the $s$- and $u$-channel resonances, 
in a treatment of the hadron structure, and in a set of experimental 
data to which the free parameters were adjusted.  
In the SLA model only one nucleon, $P_{13}(1720)$, and four hyperon, 
$S_{01}(1407)$, $S_{01}(1670)$, $P_{01}(1810)$, and $P_{11}(1660)$,   
resonances are included whereas in KM four nucleon, $S_{11}(1650)$, 
$P_{11}(1710)$, $P_{13}(1720)$, and $D_{13}(1895)$, and no hyperon 
resonances are assumed~\cite{KM}. The H2 model includes the same nucleon 
resonances as KM plus two hyperon resonances, $S_{01}(1670)$ and 
$S_{01}(1800)$~\cite{HYP03}. 
In SLA hadrons are treated as point-like objects but  
in the KM and H2 models the dipole-type hffs are included in the 
baryon-meson-baryon vertices. The models provide reasonable results 
for photon laboratory energies below 2.2 GeV, see, e.g., analysis 
of data with the SL and KM models in Ref.~\cite{ByMa}. 

In the RPR model for $K^+\Lambda$ production, the nonresonant part 
of the amplitude is modeled by exchanges of two strongly degenerate 
$K^+(494)$ and $K^{*+}(892)$ trajectories as in the Regge 
model~\cite{Regge} where the corresponding propagators can be 
assumed either with a constant (1) or rotating 
(e$^{-i\,\pi\,\alpha (t)}$, $\alpha (t)$ is the Regge trajectory) 
phase~\cite{RPR,LesleyPhD}.  
This phase ambiguity could not be removed in the version 
RPR-2007~\cite{RPR} using the standard least-squares approach 
in an analysis of high-energy data but, 
applying the Bayesian inference method in the analysis~\cite{Bayes,LesleyPhD}, 
the rotating phases were unambiguously asigned to both propagators 
in the new versions RPR-2011A and RPR-2011B~\cite{LesleyPhD}.   
In addition, to maintain gauge invariance in the RPR model, the amplitude 
includes the electric part of the proton exchange which is reggeised 
in the same way as the kaon exchange~\cite{RPR,LesleyPhD,Regge}. 
The three free parameters of background, 
the pseudo-scalar coupling constant of the $K^+$ trajectory and 
the vector and tensor coupling constants of the $K^*$ trajectory,  
are determined by fitting to photoproduction data above the resonance 
region ($\sqrt{s} = W > 2.6$ GeV)~\cite{RPR, Lesley, LesleyPhD}. 

The resonant part of the amplitude is described by exchanges 
of nucleon resonances like in the isobar model. A smooth transition 
from the resonant region into the high-energy Regge region is 
assured by strong hffs of Gaussian~\cite{RPR} or 
multidipole-Gauss~\cite{Lesley,LesleyPhD} type. In the robust analysis 
of the world's photoproduction data based on the Bayesian evidence two 
sets of nucleon resonances with highest probabilities of contributing 
to the reaction mechanism were selected from the 2048 considered model 
vaiants~\cite{Lesley}. These sets of nucleon resonances constitute 
the new versions of the Gent RPR model, RPR-2011A with eight resonances 
and RPR-2011B with five resonances~\cite{LesleyPhD}.
 
Important merit of the RPR model, besides that it describes satisfactorily 
the experimental data in the broad energy region from threshold up to 
$E_\gamma^{lab}\approx 16$ GeV ($W = 5.56$ GeV), is absence 
of the large Born contribution in the nonresonant part of the amplitude. 
Therefore, no hffs for the background are needed making 
a difference in the reaction mechanism of the RPR model and 
the isobar model with hffs. This difference appeares to be important 
for description of the cross sections at very small kaon angles as we  
will discuss in the next section.
 
\section{Discussion of results} 
\begin{figure} [htb]  
\begin{center}  
\includegraphics[width=55mm,angle=270]{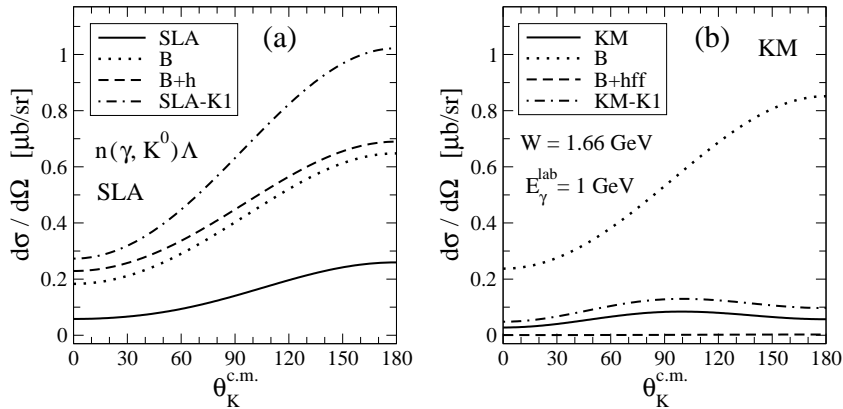}    
\end{center}  
\caption{Dynamics of the SLA (a) and KM (b) models in 
$n(\gamma,K^0)\Lambda$ at $E_\gamma^{lab}$ = 1~GeV. 
Contributions from the Born terms (B) are slightly modified by hyperon 
exchanges in SLA (B+h) but strongly reduced by hadronic form factors 
in KM (B+hff).
Solid and dash-doted lines show results of entire models and 
results without the $K_1$ exchange, respectively. The parameter 
of the $K_1^0$ exchange in SLA is from Ref.~\cite{Tsukada} 
($r_{K_1K\gamma}$ = -1.405) and in KM the original value from Ref.~\cite{KM} 
is used ($r_{K_1K\gamma}$ = -0.45).} 
\label{K0-dynamics}  
\end{figure}

The different mechanism of reducing the large contribution of the 
Born terms in the SLA and KM models plays an important role in description 
of the $n(\gamma,K^0)\Lambda$ reaction as it is demonstrated in 
Fig.~\ref{K0-dynamics}. 
In the $K^0\Lambda$ photoproduction the Born contribution reveals 
a different angular 
dependence than in the $K^+\Lambda$ channel due to absence of 
the kaon exchange and to a significant modification of the nucleon 
exchange (the electric part is missing and the anomalous magnetic 
moment changes its sign: $\mu_p = 1.79 \rightarrow \mu_n=-1.92$). 
In SLA the hyperon exchanges cannot counterbalance the large 
backward peaked Born contribution as they do in the $K^+\Lambda$ 
channel but now this effect can be provided by the $K_1$ exchange, 
see Fig.~\ref{K0-dynamics}a. The SLA model is very 
sensitive to the strength of the $K_1$ contribution as shown in 
Ref.~~\cite{K0d} 
which can be tuned to the deuteron data~\cite{Tsukada}. In the KM model, 
the Born contribution is strongly reduced by hffs as in the 
$K^+\Lambda$ channel and the $K_1$ exchange does not play 
a significant role, Fig.~\ref{K0-dynamics}b. This makes 
a notably smaller sensitivity 
to the strength of $K_1$ contribution in the KM model than 
in SLA~\cite{K0d}. The H2 model appeares to be still less 
sensitive, see Fig.5 in Ref.~\cite{K0d}.
%
%

In Fig.~\ref{K0-deuteron}, we show results of the SLA, KM2 
(our version of the Kaon-MAID model, see below), and H2 models 
for $K^0$ photoproduction on the neutron (a) and deuteron (b). 
In these models the ratio of the electromagnetic coupling 
constants for the neutral and charged modes of $K_1$, 
$r_{K_1K\gamma}$~\cite{Tsukada}, was fitted to the deuteron 
data in the low energy bin (Fig.~\ref{K0-deuteron}b) where 
the production of $\Sigma$ hyperons is negligible~\cite{Tsukada}.
The energy-averaged and angle-integrated momentum distribution 
of the $K^0\Lambda$ production on deuteron was calculated 
in PWIA~\cite{K0d}.  
The corresponding fitted values of $r_{K_1K\gamma}$ are -1.41, 0.47, 
and 7.75, for the SLA, KM2, and H2 models, respectively. A model 
dependence of this ratio, which can be related with a ratio 
of the decay widths of $K_1$~\cite{Tsukada,K0d}, is apparent. 
The large value for H2 is evoked by a very small sensitivity of 
H2 to this parameter in this energy region~\cite{K0d}.  
Note also 
the opposite sign of the ratio for the Kaon-MAID model with 
respect to the original value of -0.45~\cite{KM} used in 
Fig.~\ref{K0-dynamics}.  
%
%
\begin{figure} [htb]   
\begin{center}     
\includegraphics[width=47mm,angle=270]{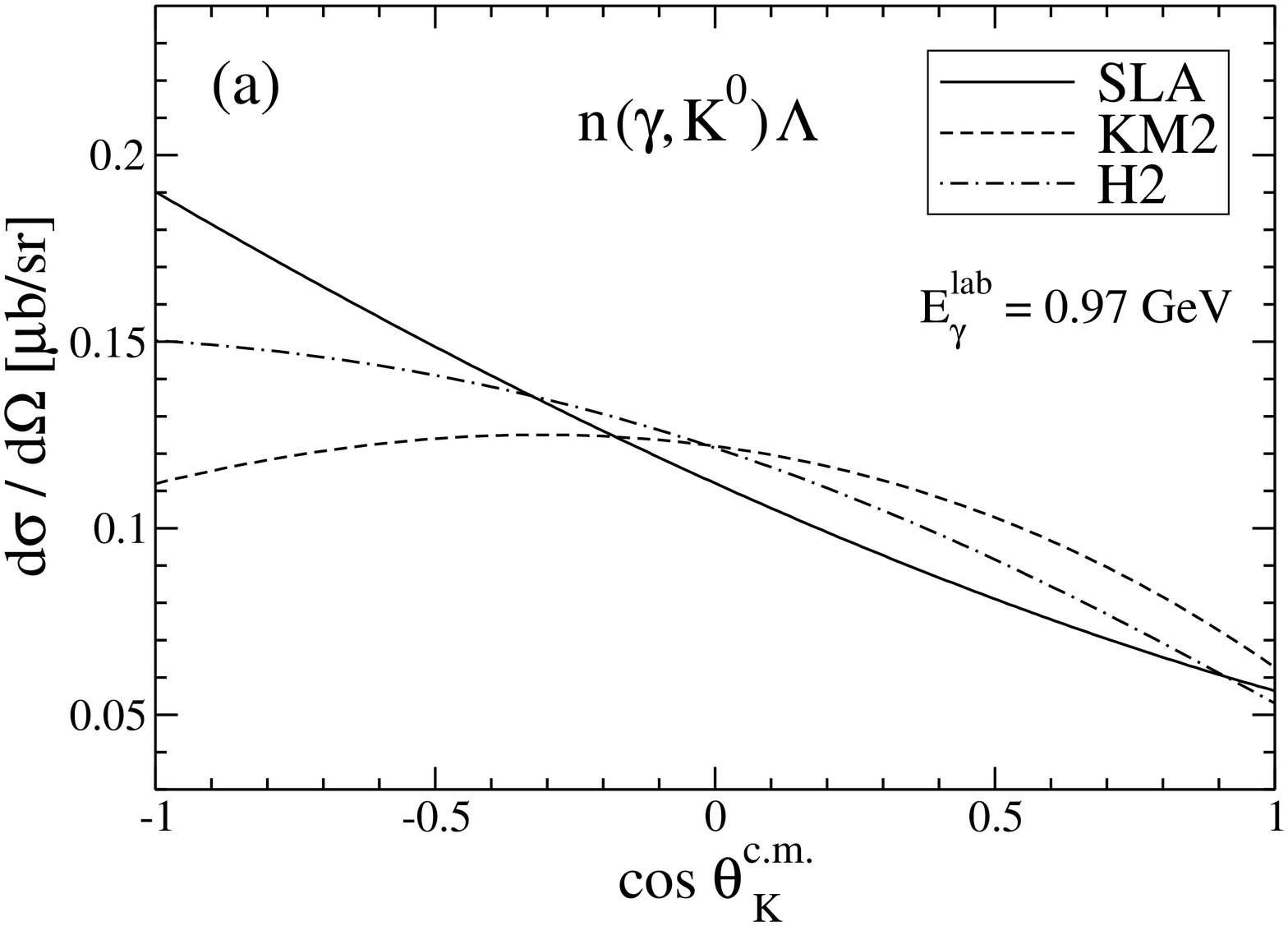}
\hspace{1mm}
\includegraphics[width=47mm,angle=270]{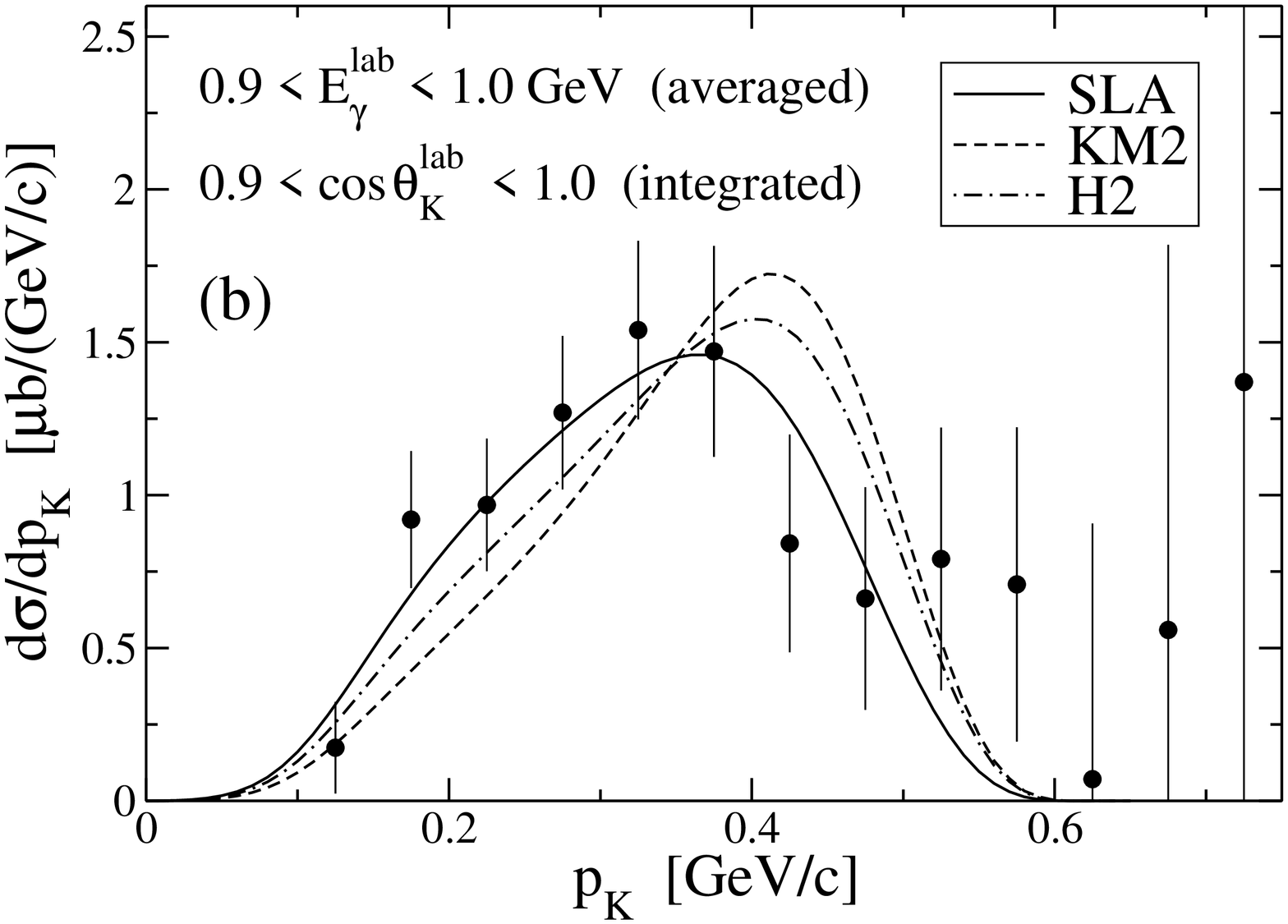}   
\end{center}   
\caption{Angular dependence of the c.m. cross sections for   
$n(\gamma,K^0)\Lambda$ is shown in (a) for the SLA, KM2, and H2 
models (see text for more details). 
PWIA calculations~\cite{K0d} of the energy-averaged 
and kaon-angle integrated momentum distributions for 
$d(\gamma,K^0$)$\Lambda p$ 
are compared with data on $d(\gamma,K^0)YN$~\cite{Tsukada} in (b). 
Contributions from the $\Sigma^0 p$ and $\Sigma^+ n$ channels are 
negligible in this energy region~\cite{Tsukada}.  
} 
\label{K0-deuteron}  
\end{figure}  

The models give different predictions for the elementary cross 
section in the backward hemisphere, see Fig.~\ref{K0-deuteron}a, 
which results in different shapes of  the momentum distributions  
for kaon laboratory momenta $0.1<p_K<0.4$~GeV/c, Fig.~\ref{K0-deuteron}b. 
The excess of the elementary cross sections in the forward 
hemisphere for the KM2 and H2 models is seen for $0.4<p_K<0.6$~GeV/c. 
Both Kaon-MAID and H2 models are not able to describe satisfactorily 
the deuteron data, even after fitting the $r_{K_1K\gamma}$ parameter, 
which can be attributed to the applied method of reduction 
of the Born terms. The SLA model fits the shape of distribution 
very well with $\chi^2/n.d.f.$ = 0.64 in contrary to 1.09 and 1.75 
for H2 and KM2, respectively.
%
%
\begin{figure} [b!]  
\begin{center}  
\includegraphics[width=49mm,angle=270]{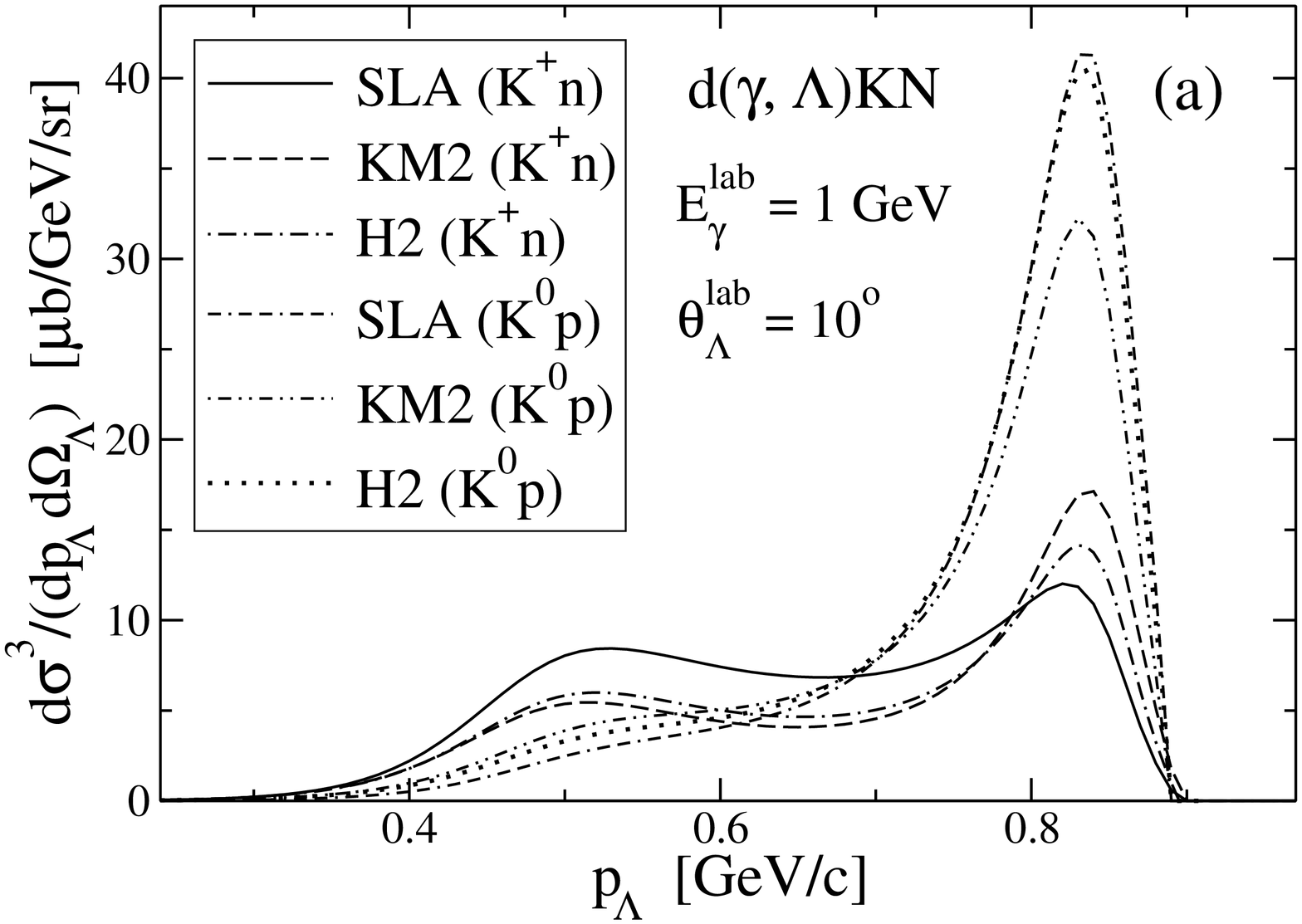} 
\hspace{2mm} 
\includegraphics[width=49mm,angle=270]{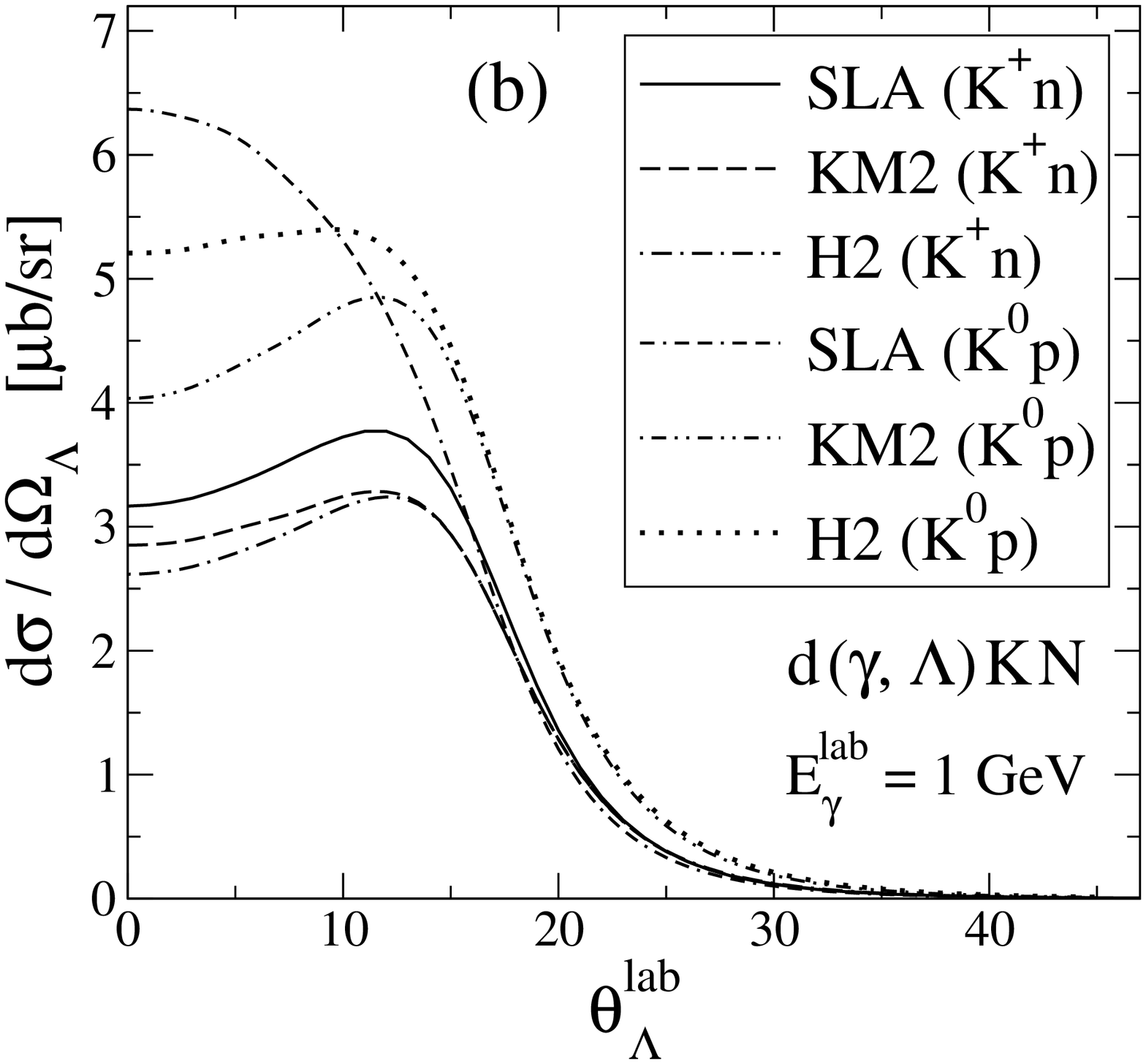}   
\end{center}  
\caption{$\Lambda$-momentum distribution at $\Lambda$ angle 
10$^\circ$ (a) and $\Lambda$-momentum-integrated angular 
distribution (b) with $K^+n$ and $K^0p$ final states in 
$d(\gamma,\Lambda)KN$ as predicted by the SLA, KM2 and H2 
models (see text for more datails).} 
\label{L-deuteron}  
\end{figure} 
   
Predictions of the SLA, KM2, and H2 models for inclusive 
photoproduction of $\Lambda$ on deuteron are given in 
Fig.~\ref{L-deuteron} for photon laboratory energy 1 GeV, displaying  
separately results with the $K^+n$ and $K^0p$ final states.
For the $\Lambda$ angle of 10$^\circ$, the models predict 
similar shapes of the momentum distribution both in the 
$K^+$ and $K^0$ production, Fig.~\ref{L-deuteron}a.  
Results of the $\Lambda$-momentum integrated cross sections in  
Fig.~\ref{L-deuteron}b are also very similar for the $K^+$  
production on proton in the whole angular range but the angular 
dependences differ for the $K^0$ production on neutron ($K^0p$ final 
state) at $\Lambda$ angles smaller than 15$^\circ$. 
All three models predict larger cross sections for 
photoproduction on the neutron than on the proton. 

%
%
\begin{figure} [b!]  
\begin{center}  
\includegraphics[width=60mm,angle=270]{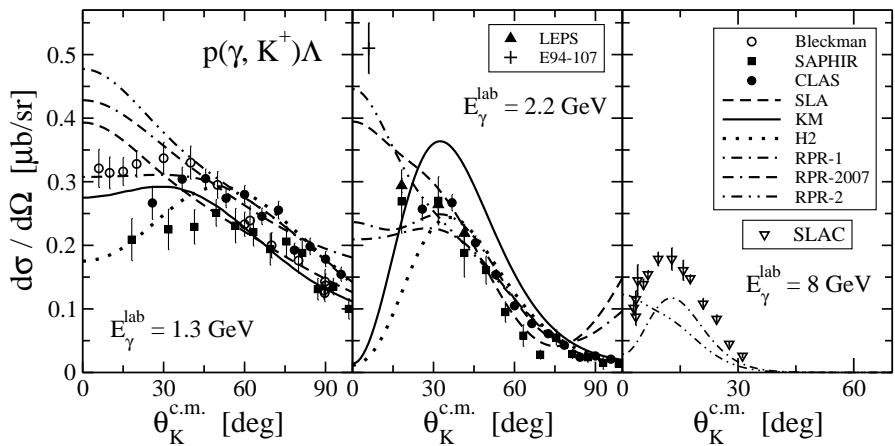}   
\end{center}  
\caption{Results of isobar and Regge-plus-resonance models 
for the cross sections in $p(\gamma,K^+)\Lambda$ are compared 
with data in the resonance region, Bleckman~\cite{Bleck}, 
SAPHIR~\cite{SAPHIR03}, CLAS~\cite{CLAS05}, and LEPS~\cite{LEPS},  
and above this region, SLAC~\cite{SLAC}. 
The Jlab Hall A data point (E94-107)~\cite{E94-107} is for 
electroproduction very near to the photoproduction point, 
$Q^2$ = 0.07~(GeV/c)$^2$, at $E_\gamma^{lab}=2.15$ GeV.} 
\label{small_ang}  
\end{figure}
In Fig.~\ref{small_ang} results of the isobar and RPR models for 
photoproduction of $K^+$ on the proton are compared with data 
for the photon laboratory energies in the resonance region, 1.3 
and 2.2 GeV, and in the Regge region at 8 GeV.
Note the problem of normalization of SLAC data~\cite{SLAC_norm} 
which, we suppose, do not affect too much their angular dependence.  
For the RPR models we adopted the version RPR-2+D$_{13}$~\cite{RPR} 
(RPR-2007) fitted to the forward-angle data and our fits RPR-1 and RPR-2. 
These new RPR models include the nucleon resonances $S_{11}(1535)$, 
$S_{11}(1650)$, $P_{11}(1710)$, $P_{13}(1720)$, and $D_{13}(1895)$ 
which were selected in the version RPR-2011B of the Gent RPR 
model~\cite{LesleyPhD} 
and which, except for the first subthreshold resonance 
$S_{11}(1535)$, are also used in the isobar models KM and H2. 
In the new RPR models the multidipole-Gauss hffs were used as in RPR-2011B. 
Both models were fitted to the cross sections from the CLAS and LEPS 
data sets; RPR-1 to data of the whole angular range and RPR-2 
only to the forward-angle data ($\theta_K < 90^\circ$). 
The models differ mainly in description of the nonresonant part 
of the amplitude. Both magnitudes and signs of the coupling 
constants of $K$ and $K^*$ trajectories differ in these models  
which appears to be important for predictions of the cross sections 
at very small kaon angles and higher energies, see Fig.~\ref{small_ang}. 

%
%
\begin{figure} [b!]  
\begin{center}  
\includegraphics[width=60mm,angle=270]{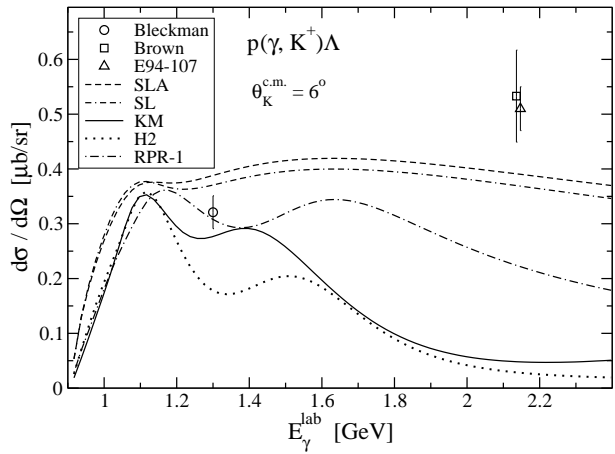}   
\end{center}  
\caption{Photon-energy dependent c.m. photoproduction 
cross sections at c.m. kaon angle 6$^\circ$ as predicted by 
isobar models and RPR-1. The data point 'Bleckman' is for 
photoproduction~\cite{Bleck} and the points 'Brown'~\cite{Brown} 
and 'E94-107'~\cite{E94-107} for electroproduction with very small 
$Q^2$. 
} 
\label{energy_dep}  
\end{figure}
In the resonance region, results of the models markedly differ  
for kaon angles smaller than 40$^\circ$, which is more apparent at 
the larger energy 2.2 GeV (Fig.~\ref{small_ang}). 
At this energy, the isobar models with hffs, KM and H2, reveal 
a strong reduction of the cross section due to 
suppression of the proton exchange by the hffs. On the contrary, 
the SLA model predicts a forward peaked cross section similarly 
as the Regge-like model RPR-2. 
A differentiation of predictions of isobar models with and without 
hffs for $E_\gamma^{lab} > 1.5$ GeV is apparent from the 
energy dependence of the cross sections for a very forward-angle 
in Fig.~\ref{energy_dep} (see also Fig.~2 in Ref.~\cite{K0d} for 
more isobar models). The lack of experimental data in this 
kinematical region, $\theta_K^{c.m.} \approx 6^\circ$, does not 
allow to test reliably the models~\cite{ByMa} and therefore to 
minimize uncertainties in the calculations of the cross sections 
for electroproduction of hypernuclei~\cite{HProd}. 

The steep angular dependence for near zero angles predicted by SLA 
and RPR-2 in Fig.~\ref{small_ang} is supported by the electroproduction 
data point E94-107 at $E_\gamma^{lab}=2.15$ GeV ($W=2.2$ GeV) 
induced by almost a real photon with $Q^2$ = 0.07 (GeV/c)$^2$~\cite{E94-107}. 
A conservative estimation of contributions from the longitudinal 
amplitudes gives for the particular kinematics the value of the transversal 
cross section (corresponding to the photoproduction cross section),
$\sigma_T\approx$ 0.38 $\mu$b/sr~\cite{E94-107}, which still favours  
the models SLA and RPR-2. The forward peaking of the cross section 
is also consistent with conclusions from the analysis of CLAS 
data~\cite{CLAS05}. The authors concluded that in the energy region 
$2.3 < W < 2.6$ GeV (2.6 GeV is the maximum energy in the experiment) 
the cross section is dominantly forward peaked 
which can be interpreted as a substantial contribution to 
the reaction mechanism by $t$-channel exchange.    
The other two Regge-like models, RPR-1 and RPR-2007, predict 
a plato at small angles in the 2.2 GeV region showing that the 
Regge-based modeling of the nonresonant part of amplitude 
can also provide other type of the angular dependence. 
Note that the SLA model is very succesful in predicting 
reasonable values of the cross sections for the electroproduction 
of hypernuclei~\cite{HYP} which points out to its realistic 
description of the elementary process at very forward kaon angles   
(dominating the hypernucleus production) and c.m. energies around 2.2 GeV. 
  
However, in the higher energy region, $E_\gamma^{lab} = 8$ GeV 
in Fig.~\ref{small_ang} ($W = 3.99$ GeV), the SLAC data~\cite{SLAC} 
reveal rather the inverse angular dependence than that observed 
in the resonance region at $E_\gamma^{lab} = 2.2$ GeV ($W = 2.24$ GeV) 
and in Ref.~\cite{CLAS05}. 
Therefore, the SLAC data, if their angular dependence will not  
change too much in a re-analysis due to the normalization, 
suggest that the RPR-1 model gives a correct angular dependence 
at very small kaon angles rather than RPR-2 which would mean 
that at 2.2 GeV a flat angular dependence (a plato) is a more 
realistic behaviour of the cross section. 
It is obvious that new good quality experimental data for 
kaon c.m. angles 0 -- 20$^\circ$ and in some energy region, e.g.,  
$2 < W < 3$ GeV, are needed to better understand the reaction 
mechanism at the very-forward-angle region.  

%
%
\begin{figure} [htb]  
\begin{center}  
\includegraphics[width=47mm,angle=270]{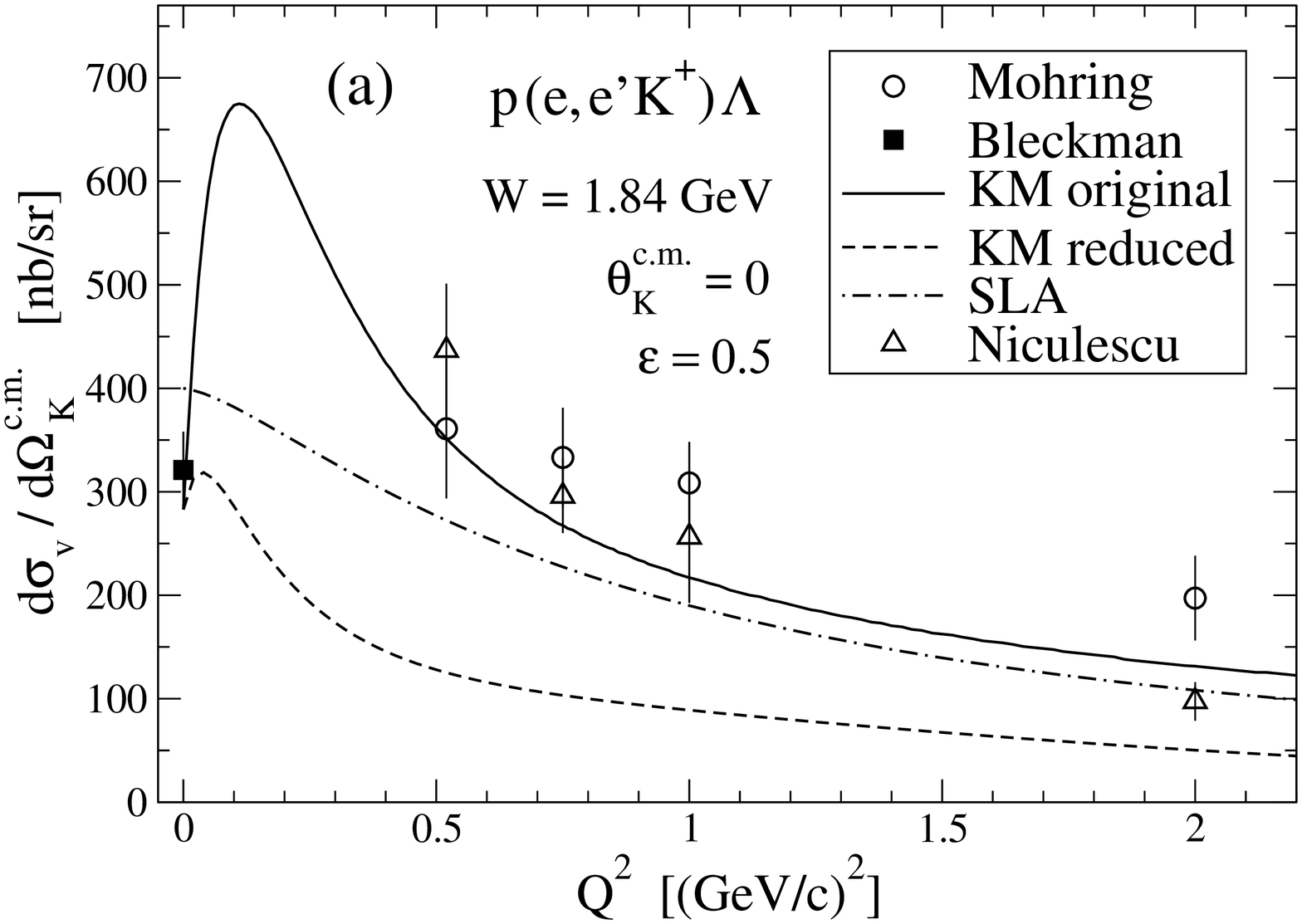} 
\hspace{1mm} 
\includegraphics[width=47mm,angle=270]{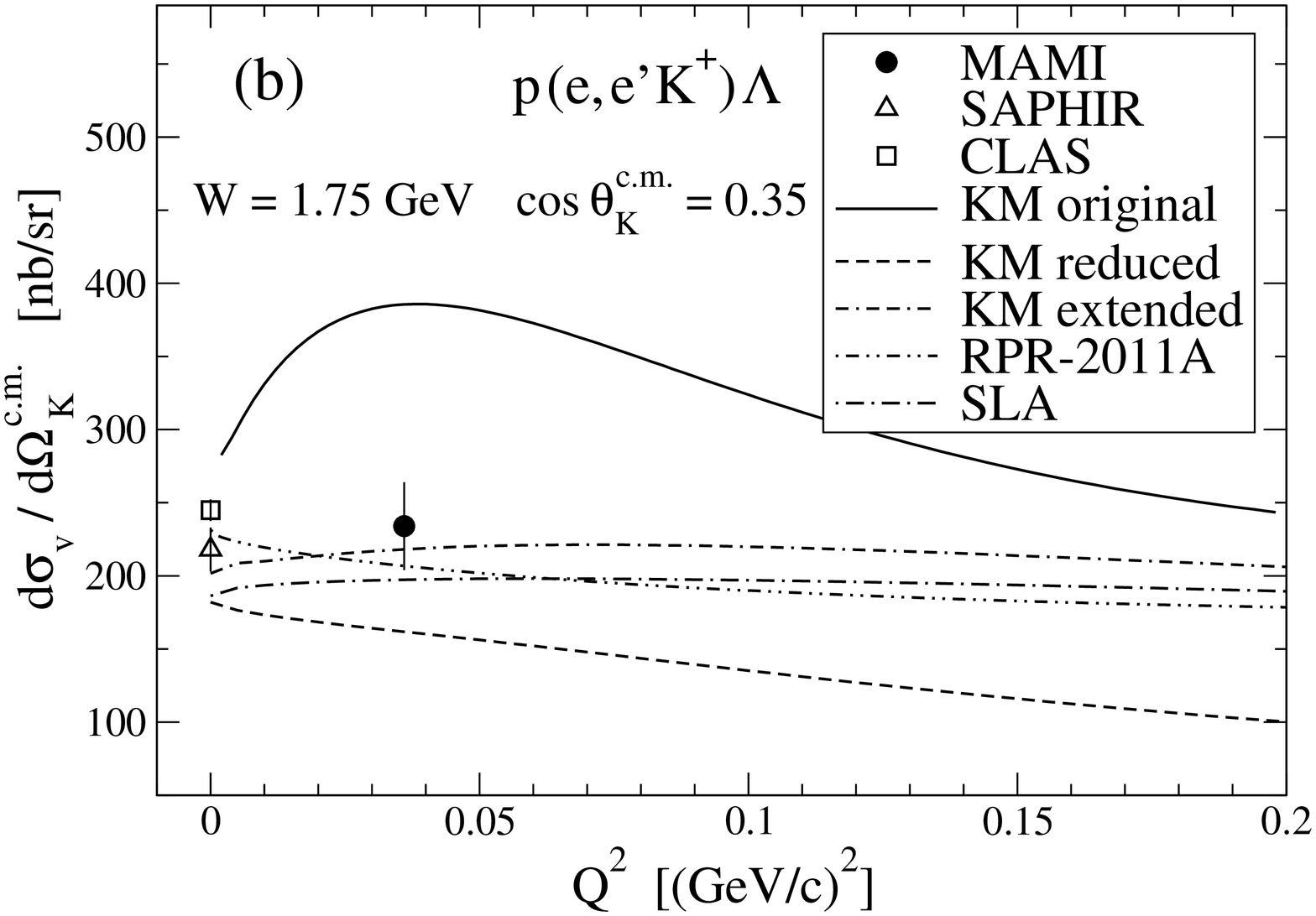}   
\end{center}  
\caption{Predictions of the isobar and Regge-plus-resonance models 
for the full unpolarized electroproduction cross section are compared 
with photoproduction data Bleckman 
(for $\theta_K^{\;c.m.} = 6^\circ$)~\cite{Bleck}, SAPHIR~\cite{SAPHIR03}, 
and CLAS~\cite{CLAS05} and with electroproduction data 
Mohring~\cite{Mohr}, Niculescu~\cite{Nicu}, and MAMI~\cite{MAMI}, 
to show a $Q^2$ dependence of the cross section for 
$0 < Q^2 < 2.2$ (GeV/c)$^2$ (a) 
and near the photoproduction point ($Q^2$=0) (b).} 
\label{small_Q2}  
\end{figure}     
In electroproduction, aside from appropriate electromagnetic 
form factors, additional possible couplings of the virtual 
photon with baryons, e.g., the ``longitudinal couplings" (LC), 
should be included in the effective Lagrangian and the 
corresponfing coupling constants should be fitted to the $Q^2$ 
dependence of electroproduction cross section. 
This was done for the Kaon-MAID model using data by Niculescu 
et al~\cite{Nicu}. The KM result, ``KM original", for the full 
cross section at the zero kaon angle, $\sigma_T +\epsilon\,\sigma_L$ 
with $\epsilon$ = 0.5, and at $W = 1.84$ GeV is shown in 
Fig.~\ref{small_Q2}a in comparison with the SLA model, 
the data by Niculescu et al, the re-analyzed data by Mohring 
et al~\cite{Mohr}, and the photoproduction data point by Bleckmann  
et al for $\theta_K^{\;c.m.} = 6^\circ$~\cite{Bleck}. 
The data by Mohring et al do not reveal such a steep $Q^2$ 
dependence as the data by Niculescu et al suggesting a smoother   
transition between the photoproduction point ($Q^2 = 0$) 
and the electroproduction data.   
The sharp bump for 0 $< Q^2 <$ 0.5 (GeV/c)$^2$ seen in the 
result of KM is modeled by strong LC as it is 
apparent from comparison with the version ``KM reduced" 
in which these couplings were removed. The SLA model, 
which does not include LC, predicts a smooth $Q^2$ dependence 
at zero kaon angle, however, overestimating the photoproduction 
data point for $\theta_K^{\;c.m.} = 6^\circ$ at this energy. 
Behavior of the full cross section near the photoproduction 
point for kaon c.m. angle about 70$^\circ$ is shown in Fig.~\ref{small_Q2}b. 
The MAMI data~\cite{MAMI} collected for a very small value 
of $Q^2$ at W = 1.75 GeV but nonzero kaon angles,  
$0.15< {\rm cos}\,\theta_K<0.65$, suggest a smooth $Q^2$ 
dependence which means that contributions  
from LC are not too big in the investigated kinematical region. 
Therefore, the models without LC can give reasonable results   
also for the electroproduction cross sections~\cite{MAMI}. 
The new version of Kaon-MAID model, ``KM extended", with 
reduced strength of LC is very well consistent with the 
new data and results from the models 
RPR-2011A~\cite{Lesley,LesleyPhD} and SLA, see also 
Refs.~\cite{MAMI} and \cite{NPA}. 

\section{Summary} 
Photoproduction of $K^0$ on the deuteron showed up to be a useful tool 
for testing the dynamics of isobar models. Data for photoproduction 
of $K^+$ on the proton (or electroproduction with a very small $Q^2$) 
at very small kaon angles and for a wide energy region are needed to 
shed light on the angular and energy behavior of the 
cross section and the dynamics of isobar models in this kinematical region. 
Recent electroproduction data for very small $Q^2$ suggest that 
longitudinal couplings of the virtual photon to baryons in the effective 
Lagrangian are not too much important in the isobar models and 
that the models without the longitudinal couplings give also reasonable 
results for the electroproduction cross sections.

\section{Acknowledgments}
P.B. wants to express a tribute and special thanks towards his friends 
and colleagues Miloslav Sotona, who passed away on 6th April, 2012, 
and Osamu Hashimoto, who passed away on 3rd February, 2012.
I am very grateful to them for the numerous stimulating and helpful 
discussions and encouragement in my work. 
We would like to dedicate this work to their memory.\\
P.B. thanks also the organizers for their kind invitation to the conference. 
This report was supported by the Grant Agency of the Czech Republic, 
grant P203/12/2126.
The work was also supported in part by the Research Infrastructure 
Integrating Activity ``Study of Strongly Interacting Matter'' 
HadronPhysics2 under the 7$^{th}$ Framework Programme of EU 
and by the Core-to-Core program of Japan Society for Promotion 
of Science.

\end{document}